\documentclass[aps,prl,twocolumn,superscriptaddress,nofootinbib]{revtex4-1}

\usepackage{graphicx, bm, amsfonts, amssymb, amsmath, appendix}
\usepackage[colorlinks=true,citecolor=blue,linkcolor=blue]{hyperref}

\newcommand{\tobs}{t_{\mathrm{obs}}}                                
\newcommand{\Tfold}{T_{\mathrm{fold}}}

\newcommand{\hego}{HeG\={o}}
\newcommand{\hogo}{HoG\={o}}

\begin{document}


\title{Rare-event trajectory ensemble analysis reveals metastable dynamical phases in lattice proteins}

\author{Antonia S. J. S. Mey}
\email[Electronic Address: ]{antonia.mey@fu-berlin.de}
\affiliation{School of Physics \& Astronomy, University of Nottingham,NG7 2RD UK}

\author{Phillip L. Geissler}
\email[Electronic Address: ]{geissler@berkeley.edu}
\affiliation{Department of Chemistry, University of California at Berkeley, Berkeley, CA 94720, USA}
\affiliation{Chemical Sciences and Physical Biosciences Division, Lawrence Berkeley National Laboratory, Berkeley CA 9420, USA}

\author{Juan P. Garrahan}
\email[Electronic Address: ]{juan.garrahan@nottingham.ac.uk}
\affiliation{School of Physics \& Astronomy, University of Nottingham,NG7 2RD UK}
\date{\today}

\begin{abstract}
We explore the dynamical large-deviations of a lattice heteropolymer model of a protein by means of path sampling of trajectories.  We uncover the existence of non-equilibrium dynamical phase-transitions in ensembles of trajectories between active and inactive dynamical phases, whose nature depends on properties of the interaction potential.   
When the full heterogeneity of interactions due to the amino-acid sequence is preserved, 
as in a fully interacting model or in a heterogeneous version of the G\={o} model
where only native interactions are considered, the transition is between the equilibrium native state and a highly native but kinetically trapped state.  In contrast, for the homogeneous G\={o} model, where there is a single native energy and the sequence plays no role, the dynamical transition is a direct consequence of the static bi-stability between unfolded and native states.  In the heterogeneous case the native-active and native-inactive states, despite their static similarity, have widely varying dynamical properties, and the transition between them occurs even in lattice proteins whose sequences are designed to make them optimal folders.  
\end{abstract}

\maketitle

In statistical mechanics, when trying to uncover the physical mechanisms behind complex emergent behavior of materials or natural systems, we very often consider highly simplified model versions of such systems in the hope that these models are simple enough to allow thorough investigation, while at the same time retaining the basic physical ingredients of the problem of interest~\cite{Chandler1987,Goldenfeld1992,Chaikin2000,Peliti2011}.  In the case of protein folding, this approach has helped shape our current understanding through theoretical and computational studies of models that discard fine details of molecular structure and/or make simplifying assumptions about the interaction energies of amino-acid residues~\citep{Shakhnovich1990,Shakhnovich2006,Rose2006}. 
One such idealized model is the representation of a heteropolymer by a self-avoiding walk on a cubic lattice, as originally proposed by G\={o}, where each occupied lattice site represents an amino acid and each edge represents an unbreakable backbone bond~\cite{Abe1981,Go1983}.  This model has been widely studied~\cite{Go1983,Socci1994,Paci2002} and has been shown to mimic the elementary aspects of protein folding, and is the system we will consider here.  

The question we address is that of the existence of highly metastable, or ``glassy'' states \cite{Weber2013}, in lattice protein models.  A well ``designed'' sequence (one that makes the protein an efficient folder) should minimize kinetic bottlenecks en route to the native state---sometimes referred to as the ``principle of minimal frustration'' in the context of natural proteins~\cite{Bryngelson1995}.  In contrast, a heteropolymer with a random or a poorly designed sequence is plagued by kinetic traps and at low temperatures gets arrested in an amorphous compact state, analogous to a glass in this context~\cite{Iben1989,Chahine2002,Gutin1998}.  On the surface at least, this would appear to indicate that in well designed protein sequences glassy states are absent.  Here we show, however, that even well designed lattice proteins, i.e.\ those with sequences that allow them to reach the desired native state efficiently, possess glass-like arrested states.  These states are thermodynamically unlikely (thus folding events are successful on average and occur fast), yet highly kinetically metastable.  We show that, in fact, dynamics takes place close to first-order coexistence between an equilibrium and ``active'' dynamical phase, and a non-equilibrium and ``inactive'' (or glassy) phase.  We also show that glass-like states can be highly native: just like in the glass problem, while active and inactive states differ markedly in their dynamics, they cannot be distinguished by simple structural measures (here degree of nativeness).  We obtain these results by studying the large-deviation properties of ensembles of dynamical trajectories via the so-called ``s-ensemble'' method \cite{Merolle2005,Garrahan2007,*Lecomte2007,Hedges2009,Jack2011} recently used to uncover dynamical phase behavior and transitions in glasses and other systems with complex dynamics.  While our results are for a highly idealized system, it is not far fetched to speculate that more detailed models of proteins will have as rich (if not richer) dynamical phase behavior as the one we report here.

\begin{figure*}[ht!]
\includegraphics[width=2\columnwidth]{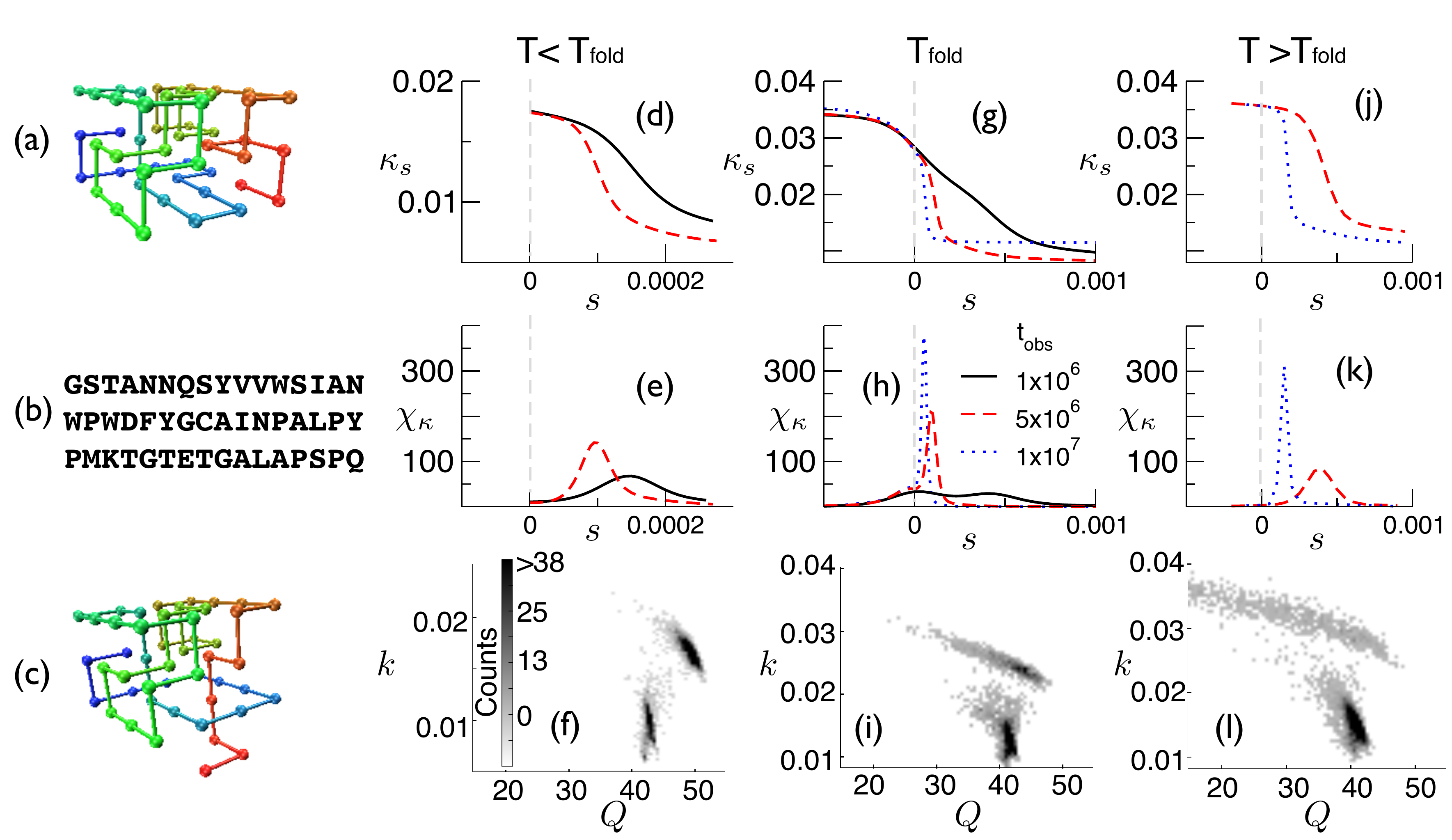}
\caption{Results of \hego~ $s$-ensemble simulations of $30000$ trajectories for different $T$ and $\tobs$ (a) native state (b) sequence (c) representative trapping state. Panels (d-f): $\kappa_s$ as a function of $s$ for temperatures $T>\Tfold=0.205$, $T=\Tfold$, and $T <\Tfold=0.175$; different curves correspond to different observational lengths $t=\tobs$ (in Monte Carlo steps), $t=1\times10^6$ (black continuous line), $t=5\times10^5$ (dashed red line) and $t=1\times10^7$ (dotted blue line); the equilibrium mean first passage time (MFPT) at $T=\Tfold$ is $t=1\times10^6$.  Panels (g-i): corresponding dynamical susceptibilities $\chi_{K}(s)$.  Panels (j-l): joint distribution of activity (per unit time and residue) $k$ and nativeness $Q$ from subsamples of $5000$ trajectories.}
\label{fig:fig1}
\end{figure*}

We study the standard lattice protein model of a self-avoiding walk on a three-dimensional cubic lattice, where each site of the walk represents an amino acid of the protein and 
each bond that connects these sites a backbone bond.  The chain is unbreakable, self-avoiding and ergodic, and allowed Monte Carlo moves maintain these properties; greater detail of the model and simulations are provided in the Supplemental Material (SM)~\cite{SM}.
With appropriately chosen interactions between residues a lattice heteropolymer displays the characteristic two-state kinetics and thermodynamics of simple proteins~\citep{Rose2006}: at high temperature the stable thermodynamic state is that of an extended and mobile chain---the ``unfolded'' state---while at low temperatures the stable state is compact and less mobile---the ``native'' state---the change of state being ``first-order like'', i.e.\ a first-order crossover (due to the finite extent of the system), whose formation is often initiated with the nucleation of a set of key contacts between residues~\cite{Gin2009,Sali1994,Garcia2006}.  As in real proteins, the folding into a specific native state is encoded in the amino acid sequence, which in turn determines the interactions.

Fig.~\ref{fig:fig1}(a) shows one of the native structures we consider: a chain of length $L=48$ with $N=57$ native contacts in its fully folded state. We analyze three possible energy functions. The first one includes all interactions between nearest neighbor residues, with an energy that depends on which two residues are involved according to the Miyazawa-Jernigan interaction matrix \cite{Miyazawa1985}---we call this the Full model; the second corresponds to considering the same interactions as in the Full model only between residues which form a {\em native contact}, i.e.\ which are nearest neighbors in the native state---we call this the heterogeneous G\={o} (\hego~) model; the third one considers only native interactions with a uniform interaction energy between native contacts---this is the homogeneous G\={o} (\hogo~) model. Results of the \hogo~ are mainly discussed in the SM~\cite{SM}. Only for the Full and \hego~ model, the sequence shown in Fig.~\ref{fig:fig1}(b) has relevance and was designed to be a fast folder to the native structure of Fig.~\ref{fig:fig1}(a)~\cite{Gin2009}. \\
The system is evolved according to standard Metropolis Monte Carlo dynamics simulations using a previously proposed ergodic move-set consisting of single and two monomer moves~\cite{Sali1994,SM}.  We are primarily interested in understanding the space-time dynamics, which is achieved by looking at rare events in the equilibrium path ensembles \cite{Merolle2005,Garrahan2007,Hedges2009}.  We denote by $X_{t}$ a trajectory, of time extension $t=\tobs$, from such an ensemble, and its corresponding path probability by $P[X_{t}]$.  In order to investigate the dynamical phase structure, we define a dynamical order parameter termed the activity $K[X_t]$, extensive in both the system size and observation time, which we will use to classify trajectories.  A convenient choice is given by the ``native activity'', that is, the total number of events in which a native contact is made or broken in a trajectory.  As in the case of glasses, activity  is the natural order parameter to explore metastability in systems displaying complex collective dynamics~\cite{Garrahan2007,Baiesi2009}.  
For an equilibrated system at a given temperature, the path ensemble $P[X_{t}]$ of trajectories $X_{t}$ can be sampled straightforwardly by generating dynamical trajectories starting from an equilibrated initial state. This is not be very efficient for exploring rare events. To explore the tails of $P[X_{t}]$ we can formally define a modified ensemble of trajectories biased by activity
\begin{equation}
P_s[X_{t}] \equiv \frac{P[X_{t}] e^{-s \, K[X_{t}]}}{Z_{t}(s)}.
\label{Ps}
\end{equation}
The parameter $s$ is a biasing ``counting'' field  conjugate to the activity $K[X_t]$~\citep{Garrahan2007}. The exponential factor in Eq.(\ref{Ps}) biases the probability of trajectories towards those which are less (more) active when $s>0$ ($s<0$) compared to the unbiased ensemble.  The normalization factor 
$Z_{t}(s) \equiv \sum_{X_{t}} P[X_{t}] e^{-s \, K[X_{t}]}$ is the moment generating function (MGF) for $K$, that is, $\langle K^{n} \rangle = (-)^{n} \partial_{s}^{n} Z_{t}(s) |_{s=0}$, and can be thought of as  
a dynamical partition function associated to the ensemble of trajectories biased with $s$. \\
In analogy with an equilibrium statistical mechanics problem, the MGF $Z_{t}(s)$ is the object of interest.  At long times it acquires a large-deviation (LD) form~\cite{Garrahan2007,Touchette2009}, $Z_{t}(s)\sim e^{t\psi(s)}$. The LD function $\psi(s)$ can be thought of as a dynamical free energy, which through a Legendre transform determines the probability $P_{t}(K)$ of observing an activity $K$ over time $t$, at long times. 
Just like the free-energy in an equilibrium problem, the analytic structure of $\psi(s)$ as a function of $s$ tells us about {\em dynamical} phases and possible phase transitions (or crossovers in the case of systems of finite extent) between them.  In particular, the scaled native activity, 
\begin{equation}
 \kappa_s \equiv (Nt)^{-1} \langle K\rangle_s = (Nt)^{-1} \sum_{X_{t}} P_{s}[X_{t}] ~ K[X_{t}], 
 \label{Ks} 
\end{equation}
(where $N$ is the number of native contacts in the fully folded state) will serve as the order parameter which allows us to distinguish between dynamical phases in the lattice heteropolymer. 
In the SM alternative choices for the activity, such as the total number of formed/broken contacts, irrespective or whether they are native or not, give equivalent results~\cite{SM}. 

The average $\langle \cdot \rangle_s$ in Eq.\ (\ref{Ks}) is over the ensemble of trajectories biased by $s$ as in Eq.\ (\ref{Ps}), which we call $s$-ensemble.  The $s$-ensemble can be probed numerically by a variation of transition path sampling (TPS) as used in~\cite{Hedges2009,Loscar2011}, in effect a Monte Carlo scheme in trajectory space that samples the distribution $P_{s}[X_{t}]$.  A trajectory $X_{t}$ can be sliced into $n$ segments of time extent $\tau = t/n$.  Each segment of the trajectory can serve as a shooting point from which part of the original trajectory is regenerated. 
For the reversible trajectories we are sampling here, the shooting direction can be forwards or backwards.  The probability of accepting a new trajectory $X_{t}^{{\rm{(new)}}}$ thus generated is given by a Metropolis criterion dependent on the change in activity, $\min[1,e^{-s(K_{\mathrm{new}}-K_{\mathrm{old}})}]$. This procedure guarantees eventual convergence to the $s$-ensemble $P_{s}[X_{t}]$ at a given temperature of the system. 
\begin{figure}[ht!]
\includegraphics[width=\columnwidth]{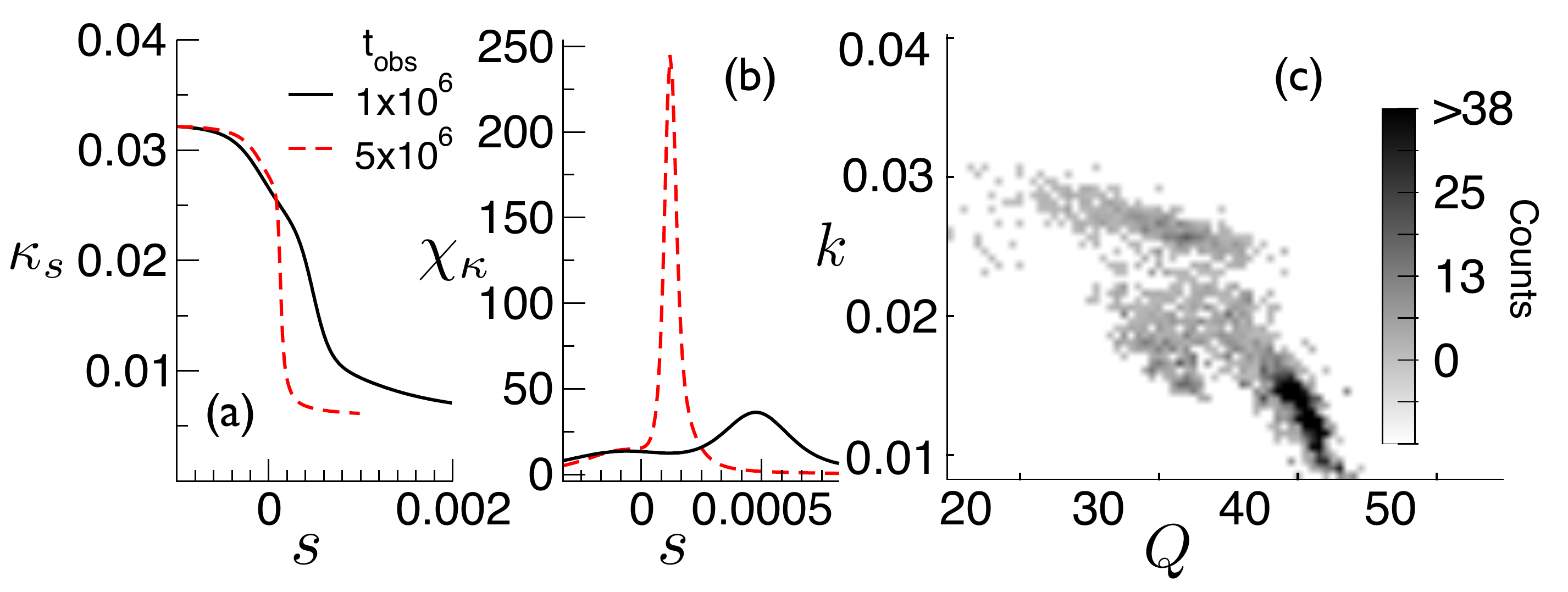}
\caption{Results of the Full model where the equilibrium ($s=0$) MFPT to the folded state is $t=2\times10^6$ at $\Tfold$(a) shows $\kappa_s$ with respect to $s$ at $\Tfold = 0.19$ averaged over $20000$ trajectories and (b) the fluctuations $\chi_{K}$ in terms of $s$. (c) Joint probability density of $k$ and $Q$ from a subsample of 5000 trajectories. }
\label{fig:fig2}
\end{figure} 

In Fig.~\ref{fig:fig1} we show results for the $s$-ensemble for the \hego~ model with the native structure and sequence of Figs.~\ref{fig:fig1} (a,b);   Figs.~\ref{fig:fig1} (d,e,f) show the average native activity $\kappa_s$, Eq.\ (\ref{Ks}), as a function of $s$ for temperatures above, at, and below the folding temperature $\Tfold = 0.19$ for this system ($\Tfold$ is defined as the temperature at which $50\%$ of the native contacts are formed on average in equilibrium).  For $s<0$ the native activity is larger than the typical one, and for $s>0$ is smaller, as expected from Eq.\ (\ref{Ps}).  But what is notable is that the change from more active to less active as a function of $s$ becomes sharp with increasing time of the trajectory; see also the dynamical susceptibilities, $\chi_{K}(s) \equiv \left( \langle K^{2} \rangle_s - \langle K \rangle^{2}_s \right)/(Nt)$, in Figs.~\ref{fig:fig1} (g,h,i). Such behavior is indicative of a first-order transition between an active/equilibrium dynamical phase, and an inactive/metastable dynamical phase.  The transition is rounded since the protein is a system of finite size in time and space.  The first-order dynamical transition in trajectories shown in Fig.~\ref{fig:fig1} is highly reminiscent of what is observed in models of glasses, where the inactive phase is associated to dynamical metastability \cite{Hedges2009,Speck2012}.

The natural question to ask is whether there is a structural signature of the active-inactive transition we observe in the \hego~ model.  An obvious structural order parameter is the ``nativeness'', i.e.\ the overall time-average of formed native contacts $Q \equiv t^{-1} \int_{0}^{t} \sum_{a} n_{a}(t') dt'$, where $a$ runs over all native contacts ($a=1,\ldots,57$ for the specific case of Fig.~\ref{fig:fig1}), and $n_{a}(t)=1,0$ indicates whether native contact $a$ is made/broken at time $t$.  Figs.~\ref{fig:fig1}(j,k,l) show the joint probability of activity (per unit time and residue) $k \equiv (Nt)^{-1}K$ and nativeness $Q$ of all trajectories in an $s$-ensemble near the critical value $s_{c}(T)$. The bimodality of the distribution is evident, as expected from the first-order nature of the transition, but what is notable is that the active/inactive basins, which differ greatly in dynamical activity, are difficult to distinguish in terms of nativeness: inactive trajectories are a consequence of a highly metastable state which is also as {\em native} as equilibrium states; a characteristic conformation associated with this inactive near-native state is shown in Fig.\ref{fig:fig1}(c).  This again is reminiscent of glasses, where the inactive glassy state is difficult to distinguish structurally from the relaxing liquid~\cite{Jack2011}.  

The existence of inactive yet highly native states seems to require the heterogeneity of interactions associated with the amino acid sequence.  In the SM we present a similar $s$-ensemble analysis for a \hogo~ model for the same structure of Fig.~\ref{fig:fig1}(a): for this model, where the sequence plays no role in the interactions, there is a straightforward linear relation between dynamical activity $k$ and structural nativeness $Q$, indicating that kinetic trapping is determined by the same structural states---unfolded and native---that determine the thermodynamics.  In contrast, the Full model for the sequence of Fig.~\ref{fig:fig1}(b), shows active-inactive transitions of similar richness as the \hego~ model, see Fig.~\ref{fig:fig2}. This is indicative of the fact for qualitative folding models sequence heterogeneity is essential, as for the case of the Full and \hego\ models~\cite{Gin2009}.

\begin{figure}[ht!]
\includegraphics[width=\columnwidth]{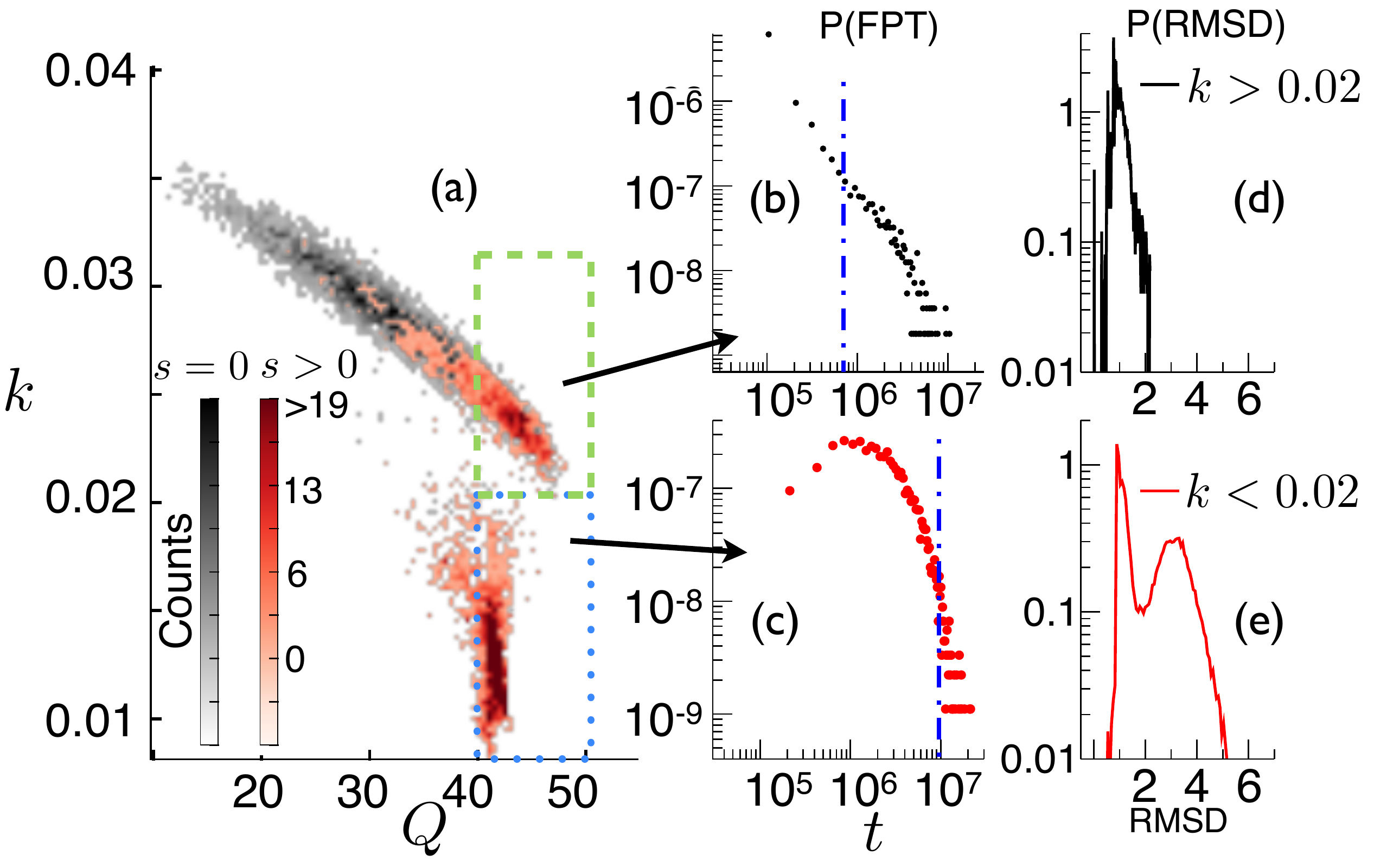}
\caption{Dynamic phase classification (a) densities of the joint distribution of $k$ and $Q$ for $s=0$ (black) and $s=1\times10^{-4}$ (red) for $\tobs=5\times10^6$ and $T=\Tfold$ of $5000$ trajectories. (b) FPT distribution of 5000 FPT trajectories for initial configurations selected from critical trajectories with $Q>40$ and $k>0.02$ and (c) for $Q>40$ and $k<0.02$. Dotted lines indicate MFPTs of the distributions. (d) Shows the RMSD distribution for initial states of $Q>40$ and $k>0.02$ and (e) with $Q>40$ and $k<0.02$. }
\label{fig:fig3}
\end{figure}
In Fig.~\ref{fig:fig3} we further explore the nature of the active and inactive dynamical phases. Fig.~\ref{fig:fig3}(a) shows
the scatter plot of the activity and nativeness $k$ and $Q$ of each of 5000 trajectories at $T=\Tfold$, for two different $s$: at $s=0$ (typical dynamics), and at $s=s_{c}\approx 10^{-4}$ [an enlarged version of Fig.~\ref{fig:fig1}(k)].  In the former case there is a clear correlation between $k$ and $Q$, while at ``space-time'' coexistence between the two dynamical phases two clear basins of very distinct activity but similar structural nativeness are found, as previously discussed. Representative configurations taken from trajectories fulfilling requirements of $Q>40$ and $k<0.02$ or $k>0.02$ respectively are used in order to investigate the dynamic nature of the highly native active or inactive trajectories. This is indicated in Fig.~\ref{fig:fig3} (a) by the respective boxes (green $k>0.02$, blue $k<0.02$). Figs.~\ref{fig:fig3}(b,c) look at the distributions of first passage times (FPT) to the fully native state starting from a typical conformation of the active or inactive phases, respectively, rapidly quenched to equilibrium ($s=0$). The inactive states not only have a much larger MFPT (indicated by the blue broken lines respectively), but the distribution is exponential, rather than stretched (as a power law, as is characteristic of near native equilibrium states in this model).
Figs.~\ref{fig:fig3}(d) and (e) show the distribution of the root mean square distance (RMSD) in the FPT trajectories. 
Active trajectories are sharply peaked around RMSD values of $1$ lattice unit, corresponding to the RMSD value of the initial structures, which is already "en route" to the native state, as seen in Fig.~\ref{fig:fig3}(d). Low activity initial structures escape trapped states by an unfolding event (RMSD peak of $3$ lattice units) followed by the folding event, see Fig.~\ref{fig:fig3}(e). This supports the idea of a long lived and metastable trapping state, prolonging folding times significantly. 

\begin{figure}[ht!]
\includegraphics[width=\columnwidth]{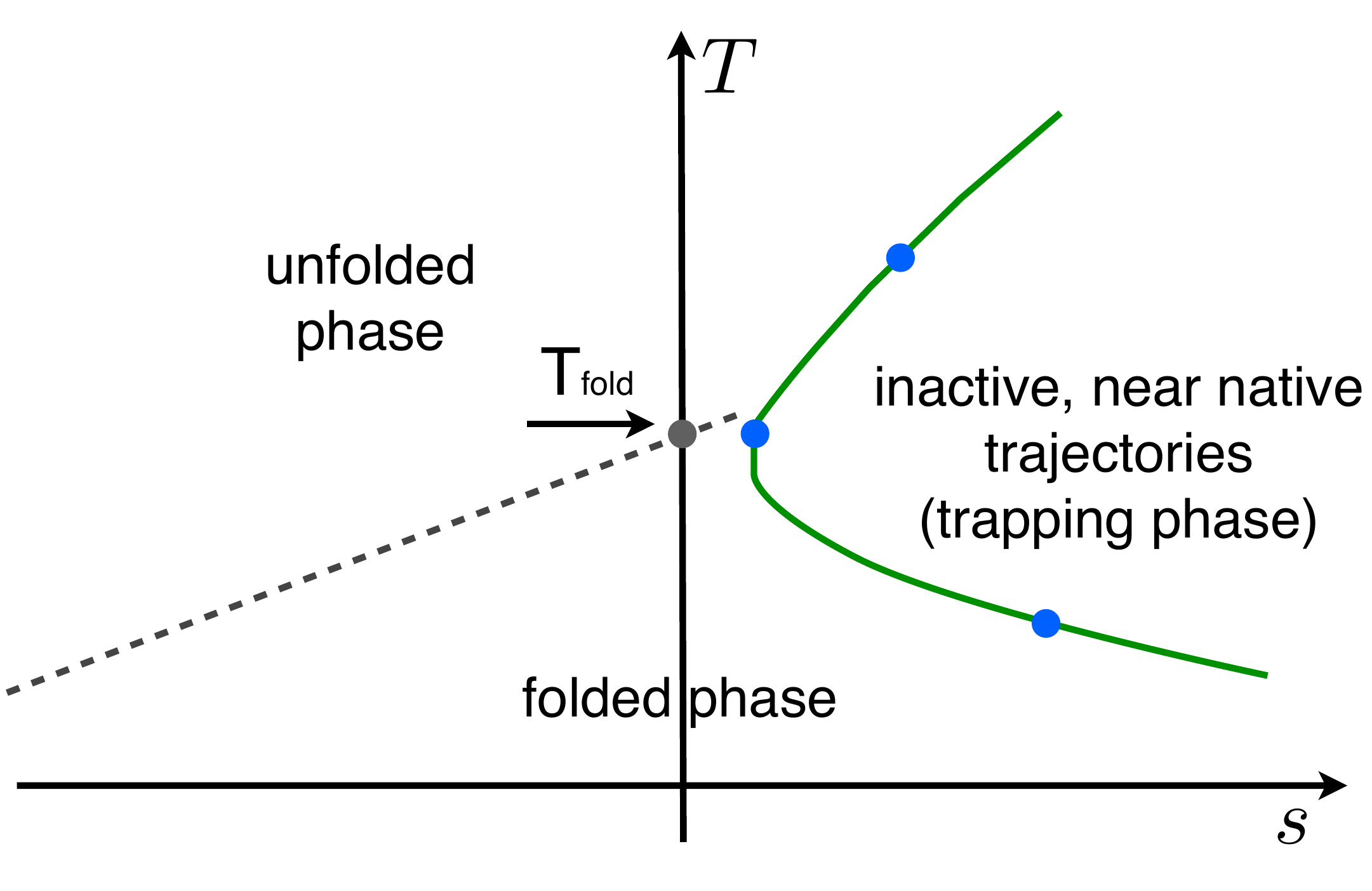}
\caption{Schematic dynamic phase diagram of the the \hego~ and Full model in the parameter space of $\{s,T\}$. First-order dynamic transition, giving rise to inactive, yet highly native trajectories, is indicated by the full green line. Blue circles along the line of the first-order transitions are taken from the results of Fig.~\ref{fig:fig1}. The dashed grey line is an extension of the thermodynamic native -- non-native transition into the $s$-space .}
\label{fig:fig4}
\end{figure}

From our simulation results, a schematic phase diagram in the parameter space of $\{T,s\}$ can be constructed as seen in Fig.~\ref{fig:fig4}.  
For all values of $T$ there is a dynamical phase corresponding to equilibrium trajectories that is smoothly connected to $s=0$.  The phase boundary of this equilibrium dynamical phase is at $s_{c}(T)$, indicated by the full green curve in Fig.~\ref{fig:fig4}. At temperatures $T > \Tfold$, equilibrium trajectories are very active as the typical states are unfolded (unfolded phase), while at $T < \Tfold$ equilibrium trajectories display a lower activity since typical states are native (folded phase).  The transition between these two regimes, indicated by the dashed grey line in Fig.~\ref{fig:fig4}, is a direct extension of the thermodynamic crossover at 
$\Tfold$ (grey circle).  At $s_{c}(T)$ there is a transition to the non-equilibrium phase of metastable, or trapped, trajectories of very low activity (trapping phase). (Three blue circles, corresponding to the three temperatures of Fig.~\ref{fig:fig1}). We surmise the phase diagram of Fig.~\ref{fig:fig4} from the analysis of the results presented in Fig.~\ref{fig:fig1}, but as we show in the SM \cite{SM}, such behavior is also valid for different sequences optimized for the same native state of Fig.~\ref{fig:fig1}(a).
Depending on the sequence and native state, the phase of inactive trapping trajectories may contain even richer phase structure, but we expect it to fundamentally be similar to that of Fig.~\ref{fig:fig4}.
 
We have shown that very simplified models of proteins, as long as interaction variability is maintained, display a complex dynamical phase structure analogous to that of glassy systems. Just like in ordinary many-body systems, the mere existence of a phase-transition to a highly inactive phase implies that rare fluctuations will play a role in the observed dynamics. 
In very recent work~\cite{Weber2013} the s-ensemble was applied to Markov state models (MSM) of real proteins. While a crossover between active and inactive phases was observed in all cases studied, no sharp transition to a glass-like state, like the one we presented here in lattice proteins, was found. This may be due to the fact that the biased distribution itself was not sampled, as only the equilibrium MSM was biased according to the $s$-ensemble with observation times no larger than the folding timescale. In contrast, our observation times were up to 10 times the MFPT of the system and explicitly sample from the biased distribution, thus observing much sharper transitions to non-native, metastable states. It can be conjectured that that fibril formation might also involve similar kinetically rare trapping events that predispose protein molecules to aggregation~\cite{Chiti2009}, which would require further extensive study of all atom models. 

\begin{acknowledgments}
ASJSM would like to thank John Chodera and Brian Gin for useful discussions and acknowledge the BESTS scholarship for funding a visit to UC Berkeley. Simulations were performed at the University of Nottingham HPC facility. 
\end{acknowledgments}

%

\clearpage
\section{Supplementary Material}
\subsection{The MC simulation setup}
As discussed in the main text, a heteropolymer of length $L$ and $N$ native contacts, represented by a self avoiding walk on a lattice, is studied. Its native state is defined by a maximally compact structure as depicted by the schematic of Fig.~\ref{fig:fig1}(a) of the main text. The choice of the native state was motivated by a series of previous studies~\cite{Gin2009, Shakhnovich1994}.
As mentioned, we chose to look at three different variants for the interaction potentials of amino acid monomers: a G\={o} interaction potential (\hogo), considering only contacts that are nearest neighbors in the native state and have a uniform interaction energy, the heterogeneous G\={o} potential (\hego), with non-uniform interaction energies between nearest neighbors in the native state and a full interaction potential (Full), allowing non-uniform interactions between all nearest neighbor amino acids.
This information can be captured by the Hamiltonian:
\begin{equation}
H=\sum_{i=1}^{L-1}\sum_{j>i}^L U(r_{ij})+\sum_{i=1}^{L-3} \sum_{j=i+3}^L N_{ij}B_{ij}\Delta(r_{ij}-a),
\end{equation}
with $r_{ij}=\vert r_i-r_j\vert$. The potential $U(r)$ restricts the walker to be self avoiding, as it takes a value of $\infty$ for $r=0$ and $0$ for any value of $r>0$. The term $B_{ij}$ is an energy interaction matrix, which is determined by the sequence of the amino acids chosen (see for example that of Fig.~\ref{fig:fig1}(b) of the main text). Interaction values for different amino acid are drawn form the model of Miyazawa and Jernigan~\cite{Miyazawa1985} and set the base value $\epsilon_0$ for the energy scale. 
In this model only nearest neighbor interactions are of interest to us and the matrix $\Delta$ holds the information of the nearest neighbor list, where $a$ is the lattice spacing (here $a=1$).
In the \hogo~ and \hego~ model $N_{ij}$ holds the information about the set of native contacts, so that $N_{ij}=1$ if the contact is present in the native state and 0 otherwise. This restriction is lifted for the Full interaction model, with all entries given by $1$. 
The Hamiltonian now allows to compute the instantaneous energy of the system at each simulation step. The simulation is a standard Monte Carlo simulation using a Metropolis acceptance criterion:
\begin{equation}
P_{\mathrm{accept}}=\min(1, \exp(-\beta\Delta H)).
\end{equation}
In order to have a set of trial moves a moveset needs to be defined. Here, it is taken to be a previously proposed moveset, consisting of single monomer and double monomer moves~\cite{Shakhnovich1994}. The single monomer moves are flips of the terminal bead or corner bead flips and the double monomer moves are "crack shaft" moves. We found a ratio of attempting $80\%$ of the time monomer moves and $20\%$ of the time crank-shaft moves to give a good sampling. 
The time observable $t$ given in the main text correspond to the incremental count of each attempt of flipping one of the beads, representing an amino acid, thus generating trajectories of any given observational time $t=\tobs$. 
Initial configurations were drawn from short equilibrium trajectories at a high temperature $T=100\epsilon_0/k_B$. The Boltzmann constant was set to $k_B=1$ in all simulations. All other temperatures will be presented unitless, defined by the Miyazawa and Jernigan interaction scale. \\

For accessing the biased s-ensemble a modified transition path sampling (TPS) strategy was employed, as briefly discussed in the main text. An initial trajectory of a predefined length of time steps ($t$) was generated. Along this trajectory with equal probability a shooting point was chosen. From which either the first or second part of the trajectory was regenerated, again with equal probability. In this way a second trajectory similar in activity $K$ to that of the first was obtained. In order to bias according to the $s$-ensemble a new trajectory was accepted according to again a Metropolis acceptance criterion, as given in the main text, by looking at the change in activity of the two different trajectories.  
\begin{equation}
P_{\mathrm{accept}}^s=\min(1, \exp(-s (K^{\mathrm{new}}-K^{\mathrm{old}}))).
\end{equation}
It should be noted, that for different simulations, discussed in the following, different numbers of trajectories are reported. In all cases a sufficient number of trajectories was generated. The variation of the number of total trajectories generated was due to available computational time and not motivated in any other way.  

\subsection{G\={o} Model simulations in and out of equilibrium}
\begin{figure}[h!]
\includegraphics[width=\columnwidth]{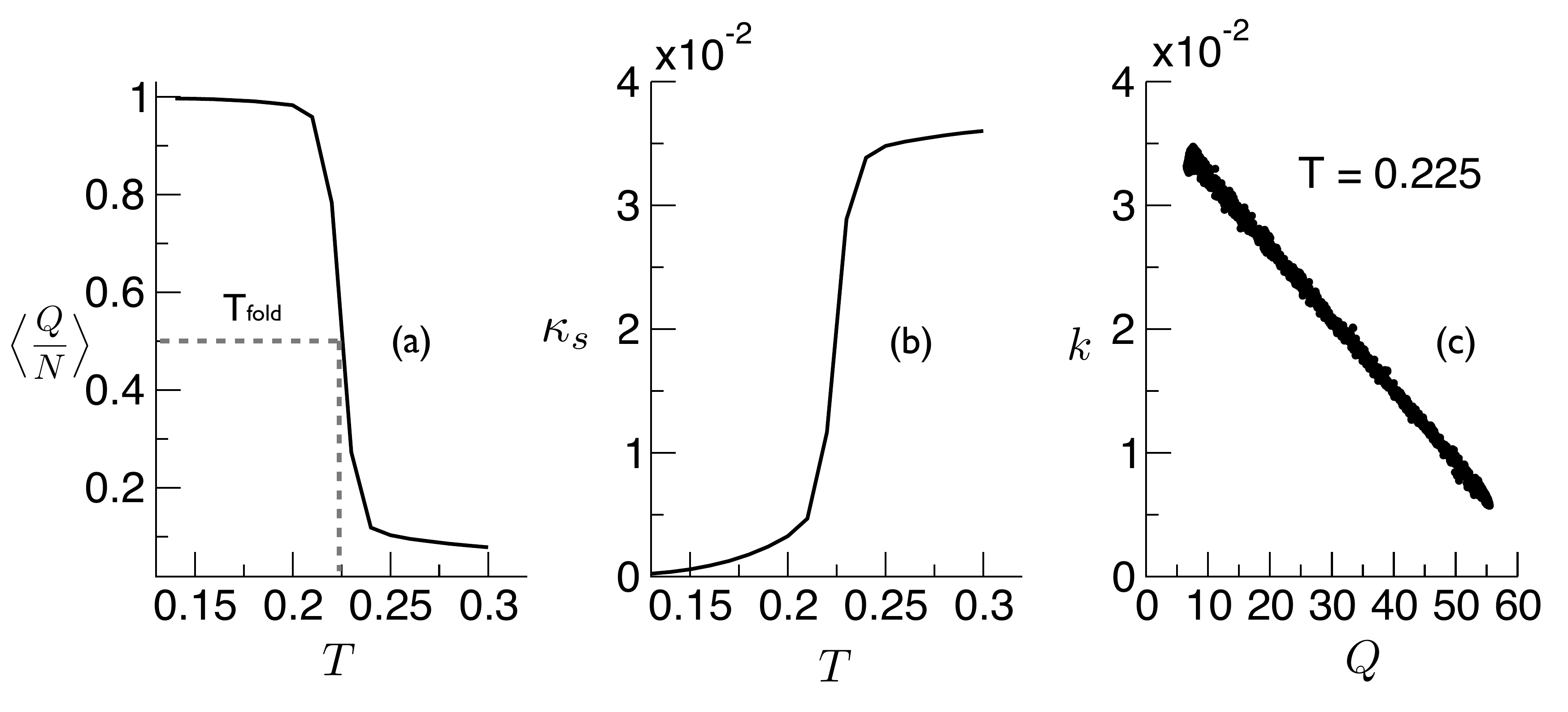}
\caption{Shows trajectory averages from $2000$ equilibrium trajectories of the \hogo~ interaction potential. Fig.~\ref{fig:figSI1}(a) shows the dependence of $Q/N$ with respect to $T$ averaged over all trajectories and in (b) the activity with respect to temperature. Fig.~\ref{fig:figSI1}(c) is the joint probability from $10000$ trajectories at $T_{\mathrm{fold}}=0.225$ of the native activity $k$ and average number of native contacts $Q$. }
\label{fig:figSI1}
\end{figure}

Initially, we looked at the behavior of a homopolymer only considering native contact interactions, i.e. the~\hogo model. For this purpose the choice of sequence was that of 48 valin amino acids, each with a valin-valin interaction energy of $E=-0.29\epsilon_0$ in reduced units, resulting in a native state energy of: $E=-16.53\epsilon_0$.
A number of $2000$ independent equilibrium simulations was conducted for $s=0$ and the average nativeness (number of native contacts over the total number of native contacts $\langle Q/N \rangle$, see the main text for definition), with respect to temperature as well as the average native activity $\langle\kappa\rangle_s$ (given by Eq.~\ref{Ks} of the main text) with respect to temperature were looked at, and displayed in Fig.~\ref{fig:figSI1}(a) and (b) respectively. Each independent trajectory was simulated for a total observation time of $t=1\times10^7$ steps. It can be seen that from Fig.~\ref{fig:figSI1}(a) the model shows also the experimentally observed sigmoidal behavior of low nativeness at high temperatures and high nativeness at low temperatures. The temperature at which $50\%$ of the native contacts are formed, $\Tfold$, can be extracted and is indicated in the Fig.~\ref{fig:figSI1}(a) (This also corresponds to the peak in the fluctuations of $\langle Q/N\rangle$). Fig.~\ref{fig:figSI1}(b) shows the dynamic behavior resulting in an inverted sigmoidal curve to that of Fig.~\ref{fig:figSI1}(a), with low activity at low temperature, corresponding to sampling native states and high activity for at high temperatures, corresponding to non-native states. 
Fig.~\ref{fig:figSI1}(c) shows the joint probability displaying a  linear dependence of the average activity ($k$) with respect to the average number of native contacts from $10000$ trajectories at $t=1\times10^7$ and at a temperature of $\Tfold=0.225$. \\
As the dynamic behavior is the subject of interest, s-ensemble TPS simulations are used to bias trajectories at $\Tfold$. No significantly different behavior was observed when considering moderate values of $s$ (i.e. $s<0.1$). 
However, the model has an intrinsic dynamic two state behavior for moderately long trajectories which will disappear when observation times of $t>1\times 10^8$ are reached. This is expected from a finite system, yet the dynamic behavior can be described as a two state dynamic cross-over and corresponds nicely to the thermodynamic first order behavior in temperature. \\

\begin{figure}[h!]
\includegraphics[width=\columnwidth]{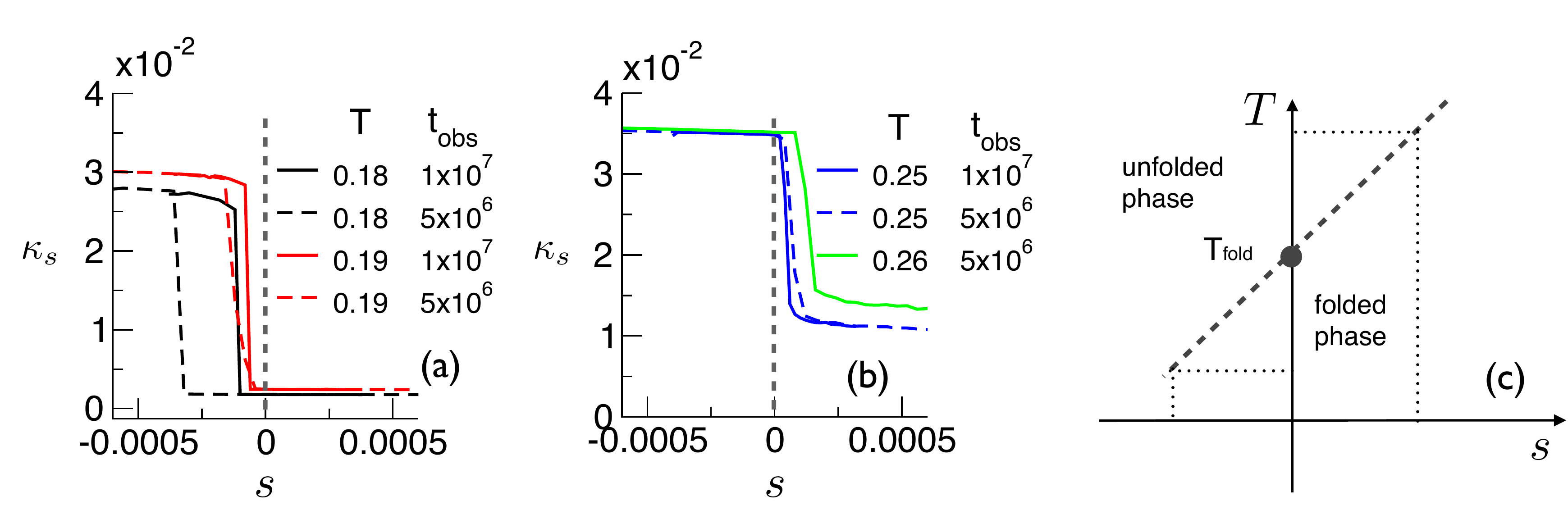}
\caption{Different temperature trajectories with the s-ensemble. Fig.~\ref{fig:figSI2}(a) shows the average native activity $\kappa_s$  with respect to the biasing field $s$ for temperature below $T_{\mathrm{fold}}$. For $s<0$ highly active trajectories can be observed. For larger $t$, $s$ approaches 0. Fig.~\ref{fig:figSI2}(b) shows again $\kappa_s$  with respect to the biasing field $s$, but for temperature above $T_{\mathrm{fold}}$. For $s>0$ the observation of inactive and mainly native trajectories is possible. For longer trajectories the biasing $s$ required is smaller. Fig.~\ref{fig:figSI2}(c) schematic of a phase diagram of active (non-native) trajectories and inactive (native) trajectories. A first order like crossover is indicted by the broken grey line, the dotted intersects are that of the observed data from a high- and a low temperature.}
\label{fig:figSI2}
\end{figure}

The dynamic equilibrium near $\Tfold$ is also accessible at lower and higher temperature, when using the s-ensemble biasing. Of course, in these cases the trajectories which are at a $50\%$ folded unfolded equilibrium are much rarer trajectories. For low temperatures the average activity of the trajectories will be low and configurations mainly sample the native and near native states. For high temperatures the opposite is the case, meaning that $\kappa_s$ is on average large and a low number of average native contacts is observed. With the biasing through $s$ a dynamic equilibrium of active (unfolded/non-native) and inactive (folded/native) trajectories can be restored even for low (Fig.~\ref{fig:figSI2}(a)) and high (Fig.~\ref{fig:figSI2}(b)) temperatures respectively as seen in Fig.~\ref{fig:figSI2}.
In Fig.~\ref{fig:figSI2}(a) results of $10000$ trajectories from $s$-ensemble biasing at two temperatures $T=0.18$ and $T=0.19$, i.e. below $\Tfold$ are shown. With an increasing temperature difference between the modelled temperature and that of $\Tfold$ the larger becomes the required bias, in order to restore mainly active trajectories as is seen in the Fig.~\ref{fig:figSI2}(a). Furthermore for larger observation times a smaller bias is needed in order to reach active trajectories, hinting at a scaling behavior with $t$. This means that for a small and finite value value of $s<0$  at temperatures of $T<\Tfold$  a dynamic equilibrium of active and inactive trajectories can be restored. 
The same applies for approaching the dynamic equilibrium at $T>\Tfold$. Now $s>0$ will restore a dynamic equilibrium. This is seen in Fig.~\ref{fig:figSI2}(b), also hinting at a scaling behaviour in observation time $t$. From these ideas a very idealized phase diagram in the parameter space of $\{T,s\}$ can be constructed, as  displayed in Fig.~\ref{fig:figSI2}(c). Here it can be seen, how with the biasing parameter $s$ the dynamic crossover can be extended to higher and lower temperatures, increasing the probability for the observation of critical trajectories. The dot indicates the merge of the dynamic and thermodynamic crossover behavior at $\Tfold$.

\subsection{\hego~ and Full equilibrium simulation results}
\begin{figure}[h!]
\includegraphics[width=\columnwidth]{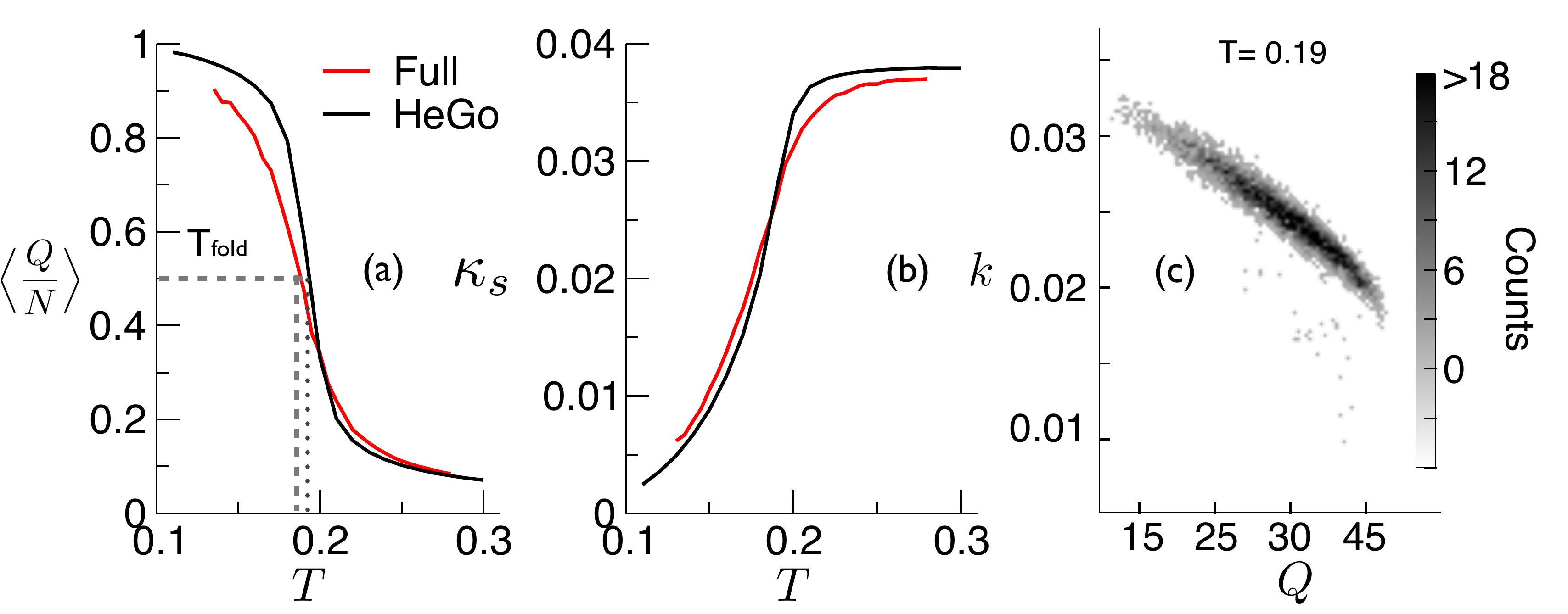}
\caption{Equilibrium behavior drawn from $1000$ trajectories of the \hego~ (black) and Full (red) interaction potential. Fig.~\ref{fig:figSI3}(a) The average normalized nativeness $\langle Q/N\rangle$ with respect to temperature is shown and (b) the average native activity $\kappa_s$ with respect to temperature is shown. Fig.~\ref{fig:figSI3}(c) joint probability from $5000$ trajectories of $k$ with respect to $Q$ at $T=0.19$ for the \hego~ potential in equilibrium.}
\label{fig:figSI3}
\end{figure}

The equilibrium behavior of the \hego~ and Full interaction model can be seen in Fig.~\ref{fig:figSI3}. Averages are taken from $1000$ independent trajectories. For both interaction potentials $\langle Q/N\rangle$ with respect to $T$ (Fig.~\ref{fig:figSI3}(a)) and equilibrium native activity $\kappa_s$ with respect to $T$(Fig.~\ref{fig:figSI3}(b)) are shown (black lines are results from the \hego~ interactions and red from the Full interaction potential). From this the folding temperature of $50\%$ native and non-native trajectories is estimated and found to be around $T=0.19$ in reduced units. This is the initial choice of temperature for the out-of-equilibrium investigation of the main text.
Fig.~\ref{fig:figSI3}(c) shows the probability density of $k$ and $Q$ from a set of $5000$ trajectories of $t=5\times10^6$ at a temperature of $T=0.19$. Most trajectories fall into a cigar-like linear dependence with a few outliers. Exactly these outliers give rise to the rich phase behavior observed in the main text. 
From the equilibrium behavior observed here, the choice of simulation temperatures and observation times used for the generation of Figs.~(\ref{fig:fig1},\ref{fig:fig2}) of the main text, was motivated. 

\subsection{\hego~ and Full $s$-ensemble simulations}
For the generation of Figs.~(\ref{fig:fig1},\ref{fig:fig2}) of the main text the following protocol was used. Simulations for different values of $s$ at three different temperatures ($T=0.175$, $T=0.19$ and $T=0.205$) were carried out. For each combination of parameter set, $30000$ trajectories were generated. For the analysis the initial $2000$ trajectories were discarded towards the equilibration of the respective value of $s$ chosen for the simulation. This method was employed for all $s$ biased trajectories. At each temperature a value for critical $s$ was identified, through the generation of histograms for each parameter set. Those with equal probability of low and high activity peaks were chosen as a set of critical trajectories and a histogram reweighting algorithm was employed in order to generate Fig.~\ref{fig:fig1} of the main text. In this way data generation could be concentrated on critical trajectories. The reweighted curves were also compared to direct estimates of average native activity $\kappa_s$ from simulations at all different values of $s$ and a good agreement was observed. 
Trajectories in the Full potential were generated in the same way and again a histogram reweighting algorithm was used on a set of critical trajectories. 

\subsection{Change of dynamic observable: general activity }
\begin{figure}[h!]
\includegraphics[width=\columnwidth]{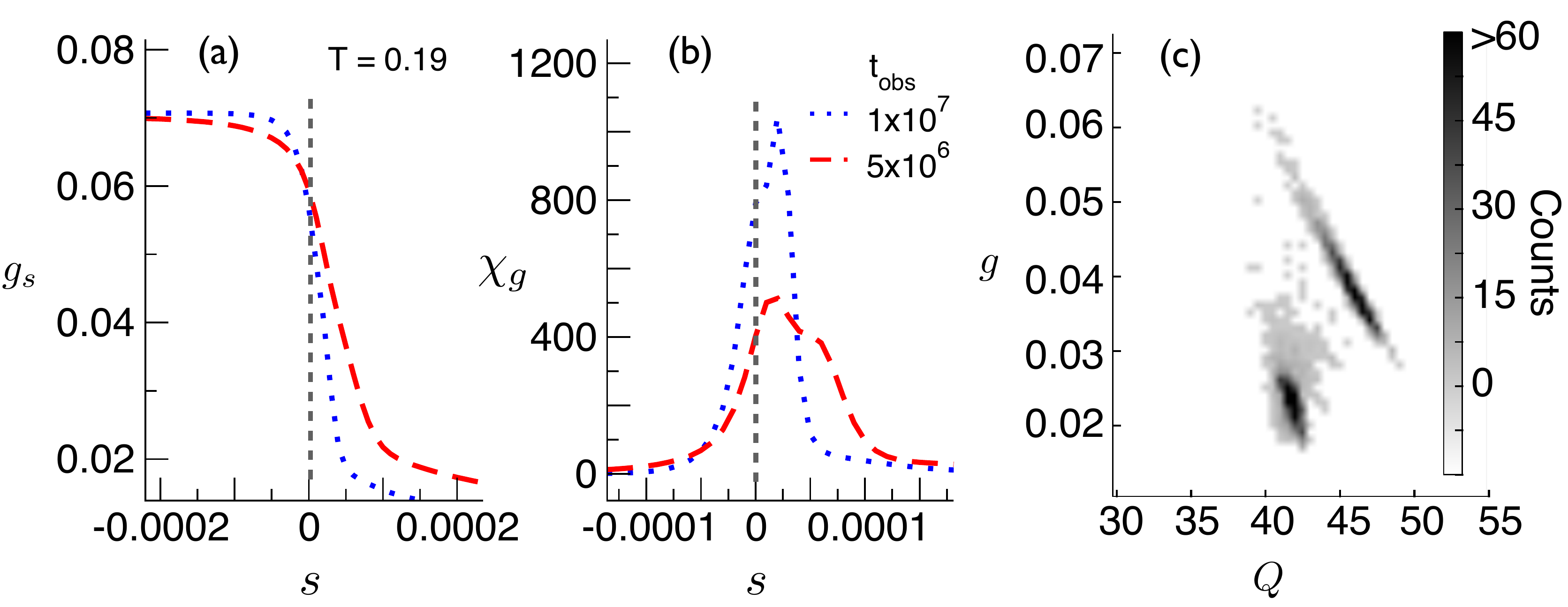}
\caption{Behavior of the \hego~ potential from averages of $30000$ trajectories with general activity $g_s$ as the order parameter at $T=0.19$. Fig.~\ref{fig:figSI4}(a) $g_s$ with respect to $s$ for $t=5\times10^6$ (red) and $t=1\times10^7$ (blue) and the fluctuations $\chi_g$ with respect to $s$ for the same $t$ in Fig.~\ref{fig:figSI4}(b). In Fig.~\ref{fig:figSI4}(a) and (b) $s=0$ is indicated by a broken grey line. Fig.~\ref{fig:figSI4}(c) shows the probability density $g$ and $Q$ from of a set of critical trajectories for $t=5\times10^6$.}
\label{fig:figSI4}
\end{figure}

As a complex dynamic phase behavior was observed for the of native activity $K$, an obvious question arises: is this complex dynamic behaviour also observed if the dynamic observable is changed? Another possible choice is that of the general activity, meaning the incremental count of any contact being broken or formed ($G$) over the whole trajectory of arbitrary length, thus the intensive general activity can be defined as:
\begin{equation}
g=\frac{G}{Nt},
\end{equation}
and averaged general activity as:
\begin{equation}
g_s=\langle G\rangle_s(Nt)^{-1}= (nt)^{-1}\sum_{X_t}P_s[X_t]G[X_t].
\end{equation}
A set of $30000$ biased trajectories for different values of $s$ at a temperature of $\Tfold=0.19$ was obtained, where $g_s$ served as the order parameter for the \hego~ interaction model.  The overall scaled activity is now larger, but again for a certain value of $s$ a sharp cross-over in the activity is observed. Taking a set of critical trajectories using a histogram reweighting method, the dependence of $g_s$ with respect to $s$ was obtained for two different observation times ($t=5\times10^6 $ (red) and $t=1\times10^7$ (blue)). See Fig.~\ref{fig:figSI4} for the results. For longer observation times the required biasing for critical behavior decreases as seen in Fig.~\ref{fig:figSI4}(a) and (b), with the fluctuations of $\chi_{g}$ with respect to $s$ are shown.

Looking at the joint probability of $g$ and $Q$ for a set of critical trajectories, the previously observed linear behavior of the equilibrium trajectories can be seen (Fig.~\ref{fig:figSI4}(c)). However, also a set of  low activity trajectories with high $Q$ is observed. These correspond to trajectories sampling mainly the highly native set of trapping states as observed for the trajectories presented in the main text. This suggests that also for the general activity observable the same complex phase behavior is present.

\subsection{Dynamic behavior of different sequences}
\begin{figure}[h!]
\includegraphics[width=\columnwidth]{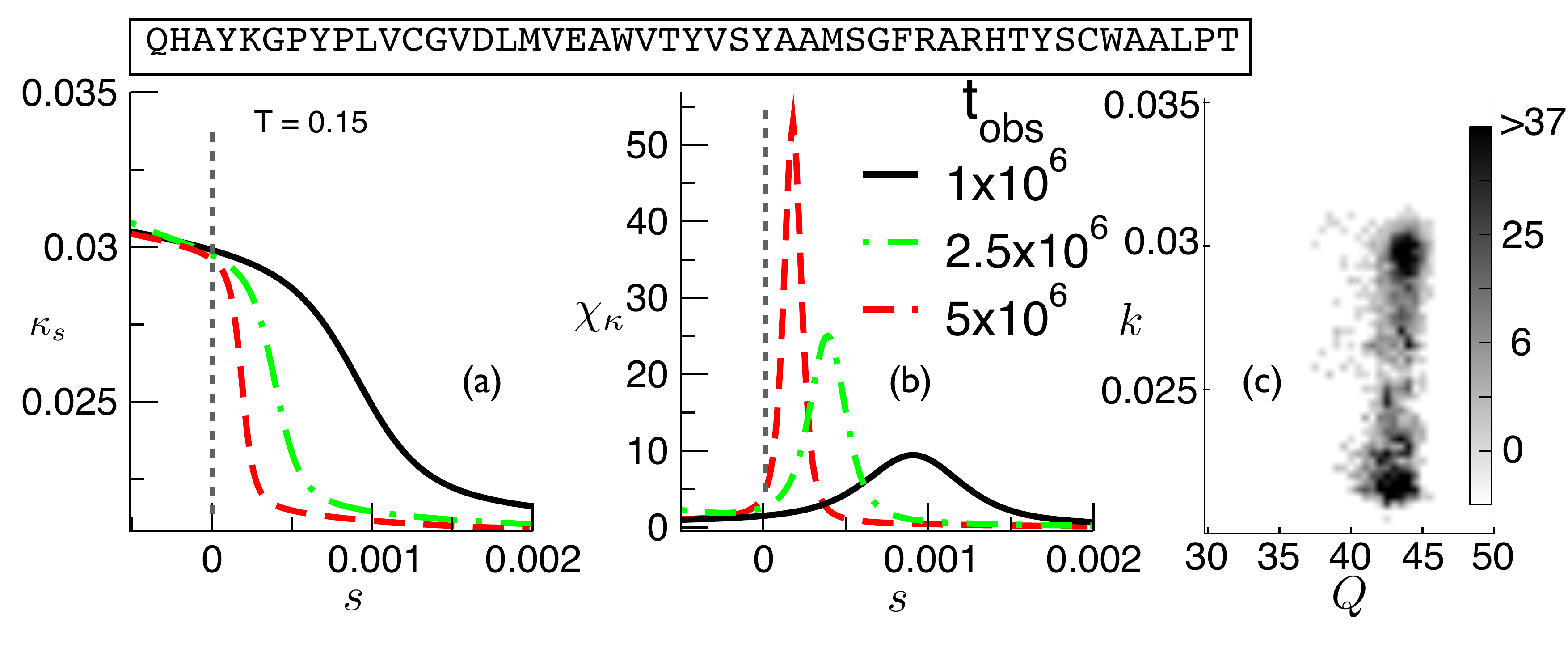}
\caption{s-ensemble behavior from averages of $10000$ trajectories for the sequence shown and native state of Fig.~\ref{fig:fig1}(a) of the main text. (a) shows the scaling of $\kappa_s$ with respect to $s$ at $T=0.15$ for different $t=1\times10^6$(black, continuous line), $t=2.5\times10^6$ (green line) and $t=5\times10^6$(red, broken line) and (b) the fluctuations $\chi_{\kappa}$ with respect to $s$ for the same $t$. In (a) and (b) $s=0$ is indicated by a broken grey line. In (c) the joint probability density for a subsample of $5000$ trajectories around a critical value of $s$ for $t=1\times10^6$, of $k$ and $Q$. }
\label{fig:figSI5}
\end{figure}

A number of different sequences, as well as chain lengths were studied in conjunction with the s-ensemble, all of which display complex dynamic behavior.  However, the nature of the phase behavior can differ drastically depending on the individual sequence and observational temperature. For example, different set of trapping states can result in different inactive trajectories, depending on the value of the biasing parameter. This rich behavior is not surprising, as for different sequences the temperature dependence can be drastically different already. In order to highlight the fact that the observation of a dynamic phase separation is more general than to one particular sequence or native structure, here we will illustrate the same observation holds for a different sequence using the \hego~ potential. For the particular sequence (as depicted in Fig.~\ref{fig:figSI5})  $\Tfold\approx0.155$ and has a native state energy of $E=12.91\epsilon_0$ in reduced units and follows a different folding pathway to the sequence presented in the main text. The mean first passage time to the folded state at a temperature near $\Tfold$ is about 5 times slower than that of the sequence presented in the main text, yet still counts as a "fast folder". Fig.~\ref{fig:figSI5}(a-c) shows the results of $s$-ensemble simulations at $T=0.15$ for different trajectory observation times $t=1\times10^6,\, 2.5\times10^6,\, 5\times10^6$. Again $\kappa_s$ serves as the activity order parameter. In Fig.~\ref{fig:figSI5}(a) the reweighted average native activity (obtained from a set $10000$ biased trajectory at a critical biasing value) is shown with respect to $s$, an increasing sharpness of the transition is observed, as the observational time is increased. In Fig.~\ref{fig:figSI5}(b) the fluctuations in $\kappa_s$ are displayed, with the peak of the fluctuations moving towards $s=0$. Fig.~\ref{fig:figSI5}(c) shows the joint probability of $k$ and $Q$ from $10000$ critical trajectories of $t = 1\times10^6$. For this sequence the observation of low activity trajectories at $s=0$ has a much higher probability than an equivalent set up for the sequence of the main text. This means, in order to achieve a dynamic equilibrium of active and inactive trajectories only small values of $s$ are needed. It should also be noted that the dependence of $k$ and $Q$ is very different to the behavior of the main sequence. There is no longer a clear linear equilibrium relationship, as seen in Fig.~\ref{fig:figSI5}(c). Another striking aspect for this sequence is, that when pushing the system even further out of equilibrium with larger values of $s$, thus raising the probability of observing very rare trajectories. In this way more dynamic complexity can be uncovered as a set of trajectories, with very low activity and a high $Q$ is found (not depicted here). These dynamic states (trajectories) may be too unlikely though to have a significant impact on equilibrium behavior. However, the existence of these states is relevant in the overall picture of the model, as it underlines a very complex dynamic phase behavior overall.    
The richness of the dynamic phases of these simple lattice models for either \hego~ or Full interaction potentials, only allows to hint at the hidden complexity to be uncovered in all atom models. 

\end{document}